\documentstyle[aps,prb,multicol]{revtex}

\title{A Thermometer for the 2D Electron Gas using 1D Thermopower} 
\author{N.~J. Appleyard, J.~T. Nicholls, M.~Y. Simmons, W.~R. Tribe and M. Pepper}  
\address{Cavendish Laboratory, Madingley Road, 
Cambridge CB3 0HE, United Kingdom} 
\date{\today} 
\begin{document}  
\maketitle 

\begin{abstract} 

We measure the temperature of a 2D electron gas in GaAs from
the thermopower of a one-dimensional ballistic constriction,
using the Mott relation to confirm the calibration 
from the electrical conductance. 
Under hot electron conditions, this technique shows
that the power loss by the electrons follows a $T^5$ dependence in the
Gruneisen-Bloch regime, 
as predicted for acoustic phonon emission
with a screened piezoelectric interaction. 
An independent measurement using 
conventional thermometry based on 
Shubnikov-de Haas oscillations gives a $T^3$ loss rate; 
we discuss reasons for this discrepancy. 

\end{abstract}

\begin{multicols}{2}

Accurate electron thermometry is needed in many aspects of low-dimensional
semiconductor physics, particularly given the increasing importance of hot
electron effects\cite{ridley91} as mesoscopic device dimensions are reduced and
electron mobility increases. Surplus heat energy in a two-dimensional electron
gas (2DEG) is rapidly shared amongst the carriers through electron-electron
interactions, and an effective electron temperature $T_e$ is established which
may be considerably higher than the crystal lattice temperature $T_l$, to which
both external thermometry and refrigeration are coupled. 
A measurement of the electron temperature is 
therefore needed to determine how an electron gas
thermalizes with its surroundings. 
Measurements of the thermoelectric response and thermal
conductivity of mesoscopic devices are also interesting in their own right, as they
provide fundamental information about electronic properties which is not
available from electrical transport measurements alone.

Although many techniques allow the measurement of bulk lattice temperatures,
the weak coupling of the electrons to their surroundings has hampered accurate
measurement of the electron temperature $T_e$. 
Previous techniques have employed the visibility of features in the electrical 
transport, notably Shubnikov-de Haas (SdH)
oscillations\cite{ridley91,ma91,manion87,kreschuk88,leadley89,balkan95}, 
but also using the zero field resistance and 
weak localization corrections.\cite{wennberg86,mittal96}
Mesoscopic effects such as Coulomb blockade
have also been applied as electron thermometers.\cite{pekola94} 
In this letter, we introduce a novel technique, where the thermoelectric response
(thermopower) of a one-dimensional constriction is 
used to measure the electron temperature. 
Self-consistent checks confirm the validity of the technique,
which we then employ to deduce the energy relaxation rate of heated electrons in a
2DEG in a GaAs/AlGaAs heterostructure, obtaining good agreement with the theory
of acoustic phonon emission in the Gruneisen-Bloch regime.

A schematic of the device is shown in Fig.~\ref{f:device}, and is similar to those
used in previous thermopower measurements.\cite{molenkamp90,dzurak93b} 
The area shaded grey shows the etched mesa 
containing a 2DEG defined at a GaAs/AlGaAs heterojunction, 
2770~\AA\  below the surface of a structure grown by
molecular beam epitaxy.\cite{thomas95} 
From measurements of the low-field SdH oscillations and 
the zero field resistivity,
the 2DEG has an electron density
$n_e=2.1\times10^{11}$~cm$^{-2}$ and mobility of 
$\mu=4.5\times10^6~$cm$^2$V$^{-1}$s$^{-1}$ at 1.5~K.
Electrons in the heating channel to the left of the device are 
heated to $T_e$ by an AC electric current $I_H$ at $f_H=85$Hz.
In a high mobility 2DEG
the thermal decay length exceeds $100~\mu$m
for $T<2$~K, and heat is efficiently conducted over the whole heating channel.
Hot electrons are prevented by the 1D constrictions A and B 
from passing into the right-hand side of the 2DEG, 
which therefore remains at the lattice temperature $T_l$. 
Each 1D constriction is formed by
depleting the 2DEG with a negative voltage $V_g$ applied to a pair of 
Schottky split-gates (shown in solid black) that are defined on the device surface, 
with lithographic length $L=0.4~\mu$m and gap width $W=0.8~\mu$m. 
On account of its thermoelectric properties,  
the temperature difference $T_e-T_l$ across a 1D 
constriction generates a potential difference, 
which is measured\cite{meas} at the harmonic frequency $2f_H$. 
As the 2DEG is heated directly, without raising the lattice temperature
$T_l$, the phonon drag contribution to the thermopower\cite{cantrell86} can be ignored and
only the contribution of electron diffusion is detected.
Figure~\ref{f:device} shows that $\Delta V$ is in fact the voltage 
measured across the two constrictions; 
one constriction is kept at fixed width as a reference, 
whilst the $V_g$ of the other is swept. 

When both charge and heat are exclusively carried by electrons,
it has been shown both for diffusive\cite{MottJones} 
and ballistic\cite{sivan86} transport that the
thermopower $S$ is related to the energy 
derivative of the conductance $G$,
\begin{equation} 
S=\left.\frac{\Delta
V}{T_e-T_l}\right|_{I=0}=-\frac{\pi^2k_B^2}{3e}(T_e+T_l)
\frac{\partial(lnG)}{\partial\mu},
\label{e:Mott}
\end{equation} 
where $\mu$ is the chemical potential of the contacts relative to the 1D subbands.
It is assumed that features in the thermopower are not subject to thermal
broadening, and that the electrons are non-interacting. 
The electrical transport properties of ballistic 
1D constrictions are well understood.\cite{BeenRev} 
The conductance is quantized at $G=i(2e^2/h)$,  
when there are $i$ transmitted 1D subbands.
As the gate voltage $V_g$ on the split-gate is made more negative,  
the constriction is narrowed and the number of 1D subbands is reduced,
and the conductance drops in steps of $2e^2/h$.
The potential in the constriction 
can be well described by a saddle point,\cite{fertig87,lmm92} 
and tunneling through the device has a 
characteristic broadening energy $\hbar\omega_x$.
As the conductance $G$ rises from $i(2e^2/h)$ to $(i+1)(2e^2/h)$ 
a peak is observed\cite{molenkamp90,dzurak93b} in the thermopower voltage $\Delta V$.
The height of the $i^{th}$ peak in $\Delta V$ is expected to be 
\begin{equation}
\Delta V_{pk}^i=-\frac{C(T_e^2-T_l^2)}{(i+\frac{1}{2})},
\label{e:Spk}
\end{equation}
where the constant $C$ depends on the intrinsic peak broadening.
In the saddle-point model, $C=\sqrt{2}\pi^3k_B^2/24e\hbar\omega_x$ 
for AC measurements in linear response.\cite{proetto91}

A simultaneous measurement is made of the four-terminal resistance $R$ of the
constriction, using a small current of $I_R=2$~nA at a different frequency
$f_R=18$~Hz. Figure~\ref{f:rawdata} shows the conductance, $G(V_g)=1/R$, of
constriction A, 
as its width is swept whilst constriction B is held fixed as a reference. 
The steps in conductance between the quantized values lead to
peaks in the measured thermopower, and we also plot the thermopower behavior
predicted from the Mott relation Eq.~\ref{e:Mott}, $d(lnG)/dV_g$. The thermal
voltage is in good agreement with this prediction, except in the region close
to pinch-off ($i=0$) which will be discussed elsewhere.

Figure~\ref{f:rescaled} shows thermopower data for constriction A at different
lattice temperatures $T_l$ and heating currents $I_H$. The traces $\Delta V(V_g)$
collapse onto a single curve when normalized by the $i=1$ peak height $\Delta
V_{pk}^1$; a similar data collapse was seen for constriction B. The dotted
line is representative of measurements taken at high lattice temperatures and
currents, where thermal broadening causes the breakdown of the Mott relation
and the use of the constriction as a thermometer. 
This broadening occurs at $T_e\approx3$~K for
constriction A, and 1.7~K for constriction B. The inset to the figure confirms
the prediction\cite{proetto91} for the height of the $i^{th}$ peak in
the thermopower voltage, $-\Delta V_{pk}^i\propto1/(i+\frac{1}{2})$, for many 1D
subbands.

From Eq.~\ref{e:Spk} we use the $i=1$ thermopower peak height to determine the
temperature $T_e$ of the hot electrons in the heating channel. 
The fitting parameter $C$ yields $\hbar\omega_x=(1.6\pm0.2)$~meV and
$(3.0\pm0.3)$~meV for constrictions A and B, respectively. Calibrating the gate
voltage in terms of energy using the method of Patel 
{\it et al.},\cite{thomas95,patel91b} 
and fitting the shape of the conductance
characteristics between the $i=1$ and $i=2$ conductance plateaux to the
saddle-point model gives the independent values $\hbar
\omega_x=(2.2\pm1.0)$~meV and $(2.0\pm0.6)$~meV. 
This provides independent confirmation that the thermopower 
follows the Mott prediction, and that our
calibration of the electron temperature scale is accurate.

Figure~\ref{f:VvsI} shows $-\Delta V_{pk}^1$ as constriction B is swept at
several lattice temperatures $T_l$, 
as a function of the power dissipated per electron $P=I_H^2R_H/n_eA$. 
The resistance $R_H=61.5~\Omega$ is the measured along the heating channel, 
which has an area $A=(800\times100)(\mu$m$)^2$. 
Electrons heated to a temperature $T_e$ 
exchange energy with the lattice at $T_l$, 
and the net power transfer is expected to follow 
\begin{equation}
P=\dot{Q}(T_e)-\dot{Q}(T_l). 
\label{e:power}
\end{equation} 
For low currents we measure $\Delta V_{pk}^1\propto P$, 
demonstrating that the thermopower
is in the linear response regime, $(T_e-T_l)<T_l$. 
At high currents, the lattice temperature becomes irrelevant to the net loss rate, 
and the data converges to $\Delta V_{pk}^1\propto P^{0.4\pm0.02}$,
suggesting that $\dot{Q}\propto~T_e^{5.0\pm0.3}$. 
A least-squares fit based on this behavior is shown as a 
dashed line in Figure~\ref{f:collapse}. 
Using the value of $C$ determined earlier, 
the thermopower peak heights collapse 
onto a single line of the form given in
Eq.~\ref{e:power} using $\dot{Q}(T)=(61\pm10) T^5 + (9\pm3) T^2$~eVs$^{-1}$
(where $T$ is in units of K); 
this form is independent of which of
the two constrictions is used as the thermometer. 
The $T^2$ term has been included to represent 
heat leakage through the Ohmic contacts of the device, 
with an effective thermal conductance $\kappa\approx 1.0$~nWK$^{-2}\cdot T$, 
equivalent to a conductance $G\approx (20~\Omega)^{-1}$. 
This term dominates at low temperatures, where cooling of
the electrons can be achieved only by thermal 
conduction through the contact wires. 
The magnitude of the thermal conduction term is of crucial importance
in device design for low temperature measurements, 
where it has proved difficult in the past 
to effectively cool a 2DEG below 50~mK.

The $T^5$ dependence of ${\dot Q}$ is expected from the stimulated emission of
acoustic phonons by hot 2D electrons. Price\cite{price82} showed that at low
temperatures, in the Gruneisen-Bloch (GB) regime, phonon emission produces
small-angle scattering of the electrons and the wavenumber-dependence of the
interaction becomes important. Coupling through a screened deformation potential
(DP) yields $\dot{Q}\propto T^7$, whereas a screened piezoelectric (PZ) coupling
gives $\dot{Q}\propto T^5$, and so PZ coupling should dominate at the
lowest temperatures in a polar material such as GaAs. Phonon dispersion
anisotropy\cite{jasiukiewicz96} reduces the prediction by a factor 0.77
to $\dot{Q}\approx270~T^5n_e^{-3/2}$ (where $T$ is in units of K and $n_e$ in
$10^{11}$cm$^{-2}$). Our measurements using the 1D thermopower for thermometry
give $\dot{Q} \approx 61~T^5$, 
in excellent agreement  with the prediction $\dot{Q}\approx 88~T^5$.  
Our measurements therefore confirm both the theories of 1D
thermopower and of 2D energy loss by phonon emission. 

The crossover from PZ to DP-dominated coupling is predicted\cite{price82} to occur at
$T_e\approx2$~K in GaAs/AlGaAs heterostructures, and so it is at first surprising
that there is no deviation from $\dot{Q}\propto T^5$ behavior at high
temperatures. However, interpolation between the equipartition
and GB regimes shows\cite{ma91} that the onset of DP-coupled scattering
merges with the demise of PZ coupling, extending the $T^5$ behavior to higher
temperatures, and hence no $T^7$ behavior is expected. We are unable to extend
our measurements of this sample to higher temperatures, to test the
prediction\cite{price82} that $\dot{Q}\propto T$ in the equipartition regime,
$T>6$~K, as thermal broadening of the thermopower peaks
invalidates Eq.~\ref{e:Spk} and our method of thermometry.

There have been few measurements of the energy relaxation from GaAs 2DEGs below
10~K in zero magnetic field. Using the 2DEG resistance for thermometry, Wennberg
{\it et al.}\cite{wennberg86} reported $\dot{Q}\propto T^5$ behavior, but with a
measured prefactor two orders of magnitude larger than expected. In
contrast, a photoconductivity measurement by Verevkin {\it et
al.}\cite{verevkin96} gave agreement with theory at the onset of the GB regime.
Recently, Mittal {\it et al.}\cite{mittal96} have measured both the $B_{\perp}=0$
resistance and the weak localization correction, and reported agreement with
Price's model around $T_e=0.2$~K, though the data presented are insufficient to
establish the full temperature dependence.

In a number
of experiments\cite{ridley91,ma91,manion87,kreschuk88,leadley89,balkan95}, the
electron temperature $T_e$ has been deduced from the damping of SdH
oscillations when the electrons are heated by varying $T_l$ or a
heating current $I_H$. These have shown that the power loss rate $\dot{Q}$
varies as $T^3$ at low temperatures, and a low-field measurement of our sample
by this technique also shows 
$\dot{Q}=I_H^2{\bar R}(B_{\perp})/N_e\propto T_e^3$, 
as shown in Fig.~\ref{f:collapse}. 
Whilst some authors have attributed
this to the crossover between the GB and equipartition regimes, our data show
that the two thermometry techniques yield different temperatures under the same
heating conditions. 
We propose, therefore, that the energy relaxation mechanism
at finite field differs from that at $B_{\perp}=$0, as the
electron wavefunctions are those appropriate to Landau levels, and the
momentum transfer in the plane is restricted to $\sim \hbar/l_c$, where $l_c$
is the cyclotron length.\cite{mckitterick94} 
Scattering is then restricted to
phonons in a narrow cone perpendicular to the 2DEG, resulting in a weaker
temperature dependence of the energy relaxation rate. 
We note that Chow {\it et al.}\cite{chow96} have observed 
$\dot{Q}(T,B_{\perp})\approx 800T^5n^{-3/2}$
using the SdH technique for $T_e<700$~mK at $B_{\perp}\approx0.14$~T, 
where the thermal cut-off of available phonon energies 
occurs before the momentum cut-off
comes into effect. 
There is some evidence of a crossover at $T_e\sim 0.5$~K
both in their data and our own, and this is an area which merits further
investigation. 

In conclusion, we present measurements of the thermopower of a 1D constriction,
including the linear response regime, that validate the single-particle model both
by comparison to simultaneous resistance measurements through the Mott
relation, and using independently obtained values for 1D subband broadening. We
have demonstrated a new method of electron thermometry, with potential for
application to the study of non-equilibrium effects in low-dimensional semiconductor
structures. This also paves the way for accurate measurements of the thermal
conductance of a 1D system and a test of theoretical
predictions\cite{kane96,fazio98} that the Wiedemann-Franz law will be violated
by interacting electrons. The measured energy relaxation rate from the 2DEG
shows unequivocally that power losses in the absence of a magnetic field scale
as $T^5$, in good agreement with the theory of coupling to acoustic phonons,
whereas the theory is no longer valid in the presence of a small
magnetic field. 

We thank the Engineering and Physical Sciences Research Council (UK) for
supporting this work, and JTN acknowledges an Advanced EPSRC Fellowship. We
thank Dr A.S.~Dzurak for useful discussions. 


\begin{thebibliography}{10}
\bibitem{meas}
This signal is sensitive only to thermal effects, which vary as $I_H^2$.
\end{thebibliography}

\end{multicols}
\newpage 

\begin{figure}
\caption{Schematic of the device and measurement circuit. The etched mesa,
shown in grey, consists of a heating channel and two voltage probes, where
the two 1D constrictions are defined. The four-terminal resistance $R$ is
measured simultaneously with the thermopower $S$, but at a different frequency.  
Magnified view: the two pairs of split-gates defining the constrictions A and B
are shown in solid black.}
\label{f:device}
\end{figure}  

\begin{figure}
\caption{Experimental traces of the conductance $G$ (derived from $R$) and the
thermopower voltage $-\Delta V$ from constriction A, using a heating current
$I_H=1.5~\mu A$ at a lattice temperature $T_l =305$~mK, so that $T_e \approx 600~$mK. 
The dashed line shows the predicted thermopower signal $-\Delta
V(V_g)\sim d(ln G)/dV_g$ from the Mott relation (Eq.~\ref{e:Mott}).} 
\label{f:rawdata} 
\end{figure}  

\begin{figure}
\caption{Collapse of nineteen traces of the thermopower
for constriction A, for lattice temperatures 0.3~K$<T_l<$1.35~K, 
and heating currents $0.7~\mu$A$<I_H<20~\mu$A. 
Each trace has been divided by the $i=1$
peak height $-\Delta V_{pk}^1$, which varies from $0.1-4.7~\mu$V.
The dotted trace, for which $T_l=1.39$~K and $I_H=20~\mu$A, shows the effect
of thermal broadening; such data have been excluded from subsequent analysis. 
Inset: The thermopower peak heights $-\Delta V_{pk}^i$ 
follow the predictions of Eq.~\ref{e:Spk}, which is shown as a solid line.}
\label{f:rescaled}
\end{figure}  

\begin{figure}
\caption{The $i=1$ thermopower peak height of constriction B as a function of
dissipated power $P$, measured at different lattice temperatures $T_l$. Linear
response is achieved even at the lowest temperatures. The solid lines show the
fit $\dot{Q}=61~T^5 + 9~T^2$~eVs$^{-1}$ (deduced from Fig.~\ref{f:collapse}).} 
\label{f:VvsI}
\end{figure}

\begin{figure}
\caption{Universal behaviour of $\dot{Q}$ from thermopower data for the two 1D
constrictions. The rms electron temperature is calculated from the
$i=1$ thermopower peak height $\Delta V_{pk}^1$. The solid curve shows
$\dot{Q}(T_e)$ and the points $\dot{Q}(T_l)+P$ 
are calculated from the heating current $I_H$. 
The dashed line is the theoretical prediction $\dot{Q}=88~T^5$,
with no fitting  parameters. Electron temperatures for the lower data
set were obtained from the amplitude of SdH oscillations for $T_e>T_l=305$~mK,
and were compared to the dependence of the low-current amplitude on $T_l$.
The dot-dash line is the fit $\dot{Q}=14~T^3$. }
\label{f:collapse}
\end{figure}

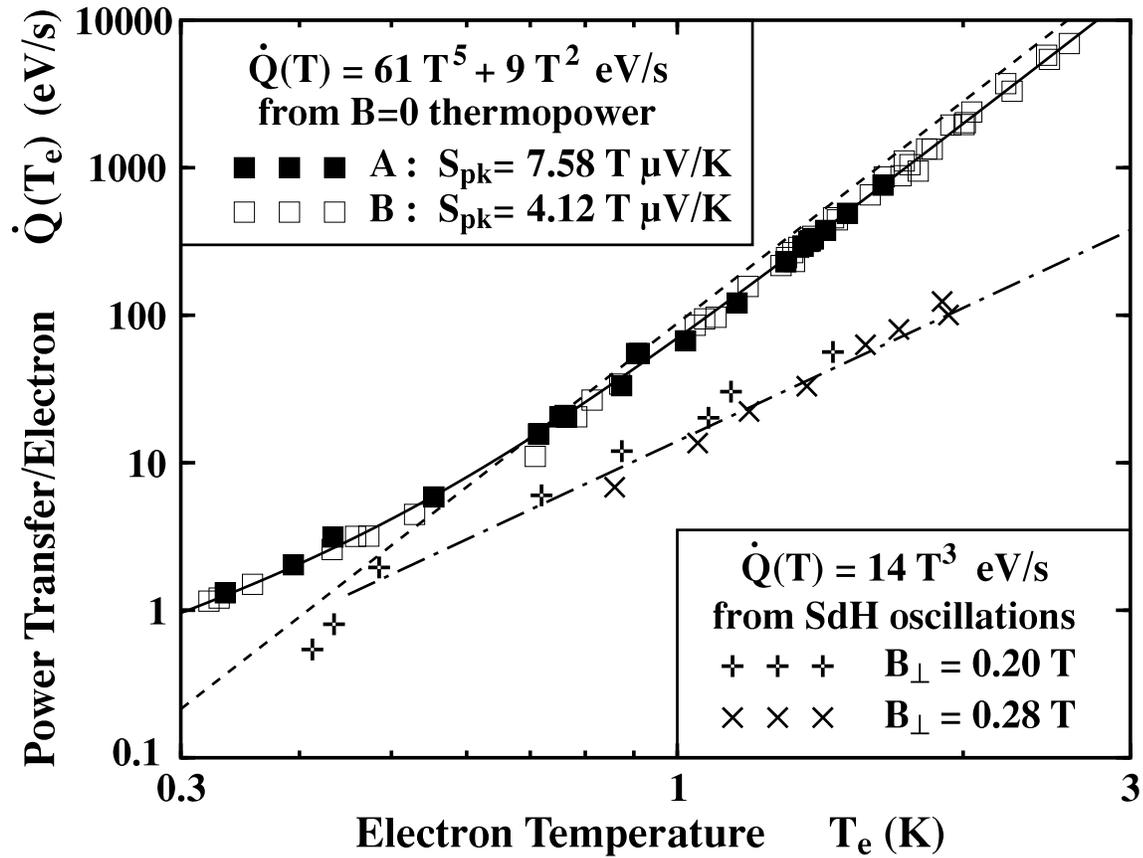
\end{document}